# Engineering Grain Boundary Commensurability for Ferroelectric Stabilization in Hafnia-Based Films

Chuqiao Shi[1#], Xinyan Li[1,2#*], Xing He[3], Kaiji Zhao[4], Jesse Schimpf[2,5], Akash Surampalli[1,2], Adan Mireles[1,10], Sergio Puebla[5], Yi Jiang[6], Ramamoorthy Ramesh[1,2,5], Xiaofeng Qian[4,8,9], Lane W. Martin[1,2,7], Andrew M. Rappe[3], Yimo Han[1,2,10*]

[1] Department of Materials Science and NanoEngineering, Rice University, Houston, TX 77005, USA

[2] Rice Advanced Materials Institute, Rice University, Houston, TX 77005, USA

[3] Department of Chemistry, University of Pennsylvania, Philadelphia, PA 19104, USA

[4] Department of Materials Science and Engineering, Texas A&M University, College Station, TX 77843, USA

[5] Department of Materials Science and Engineering, University of California, Berkeley, CA 94720, USA

[6] Advanced Photon Source, Argonne National Laboratory, Lemont, IL 60439, USA

[7] Departments of Chemistry and Physics and Astronomy, Rice University, Houston, TX 77005, USA

[8] Department of Physics and Astronomy, Texas A&M University, College Station, TX 77843, USA

[9] Department of Electrical and Computer Engineering, Texas A&M University, College Station, TX 77843, USA

[10] Smalley-Curl Institute, Rice University, Houston, TX 77005, USA

[#] These authors contributed equally: Chuqiao Shi, Xinyan Li.

[*] To whom correspondence should be addressed:

Yimo Han (Email: yimo.han@rice.edu), Xinyan Li (Email: lxy@rice.edu)

## Abstract

Grain boundaries fundamentally dictate the macroscopic properties of polycrystalline materials by breaking long-range symmetry. In ferroelectrics, these structural discontinuities are conventionally considered as detrimental features that induce depolarization fields and accumulate defects, thereby suppressing or pinning local polarization. Here, we demonstrate that the commensurability of grain boundaries inherently governs the polar-orthorhombic phase stability in polycrystalline $Hf_{0.5}Zr_{0.5}O_2$ (HZO) thin films. Using depth-resolved multislice electron ptychography, we map the three-dimensional (3D) phase distribution across diverse boundaries with atomic resolution, revealing that highly commensurate grain boundaries effectively suppress the nonpolar-tetragonal phase by compensating for lattice mismatch while mitigating geometric frustration. Monte Carlo simulations uncover the atomistic and energetic origins of polar-phase stabilization, showing that commensurate grain boundaries favor the polar-orthorhombic phase more strongly than other grain boundaries. Guided by this principle, we epitaxially engineer HZO thin films to promote the preferential formation of commensurate grain boundaries, resulting in an approximately 60% enhancement in the remanent polarization of ferroelectric devices. These findings provide direct 3D experimental and theoretical evidence that establishes grain boundary commensurability as a critical degree of freedom for designing functional interfaces, guiding the construction of novel ceramics and polycrystalline ferroelectrics.

## Introduction

In polycrystalline materials, interfaces separating crystalline grains disrupt long-range translational symmetry and provide a rich landscape for emergent properties[1–3] and structural dynamics[4,5]. Such grain boundaries are pivotal in ferroelectrics, where mechanical and electrostatic discontinuities are traditionally considered to suppress the long-range order of electric dipoles, thereby destabilizing ferroelectricity. This phenomenon, often described by the "dead-layer" model, becomes increasingly pronounced at the nanoscale[6], exerting a profound impact on device performance as functional thin films scale down to reduced thicknesses.

Standing at the forefront of this scaling challenge are hafnia-based ultrathin films[7–11] which represent a prototypical polycrystalline system[12,13]. While highly desirable for their exceptional silicon compatibility and seamless integration into modern microelectronics, these materials introduce a thermodynamic landscape that is fundamentally different from classic perovskites[14]. Specifically, ferroelectricity in hafnia does not originate from a stable ground state but rather hinges on kinetic stabilization of a metastable polar-orthorhombic phase (O phase, $Pca2_1$) relative to competing nonpolar monoclinic and tetragonal polymorphs (M phase, $P2_1/c$; T phase, $P42/nmc$)[15]. Consequently, grain boundaries in hafnia-based films are not merely structural defects that induce depolarization, but rather intrinsic structural motifs that critically modulate the phase competition and macroscopic functionality.

The structural complexity of hafnia-based thin films is dictated by their fabrication techniques. Current growth methods, such as atomic-layer deposition (ALD) and pulsed-laser deposition (PLD), enable precise control of film thickness and stoichiometry, yet the resulting microstructures are inherently polycrystalline with nanoscale grain size[12,13]. Unlike single-crystalline perovskite ferroelectrics[16], the nanocrystalline grains in hafnia-based films exhibit random orientations when grown by ALD[17] or form preferred crystallographic textures via PLD[18–20]. This inherently results in complex three-dimensional (3D) interfacial networks, including both intragranular *domain* boundaries[11,21–23] and intergranular *grain* boundaries[24] (**Fig. 1a**). Prior studies suggest that these boundaries modulate the thermodynamic stability of the metastable polar-O phase[25–31]. However, the microscopic understanding of both domain and grain boundaries remains critically overlooked, forestalling insightful guidance of active boundary engineering strategies for polar-phase stabilization and device applications.

In this work, we reveal the precise 3D phase competition between the polar-O and nonpolar-T phases across diverse boundary environments using multislice electron ptychography (MEP)[32]. We find that while domain boundaries and typical incommensurate grain boundaries favor the undesirable formation of "V"-shaped interfacial nonpolar-T phase regions (**Fig. 1b**), commensurate grain boundaries in polycrystalline hafnia-based thin films unexpectedly stabilize the polar-O phase (**Fig. 1c**). This stabilization fundamentally arises through a unique mechanism that simultaneously accommodates ferroelastic mismatch and geometric frustration, maintaining intrinsically lower grain boundary energies for the polar phase and thermodynamically prohibiting the competing nonpolar phase. Guided by these microscopic and thermodynamic insights, we deliberately engineer highly commensurate grain boundaries to enhance macroscopic device performance. These findings resolve the underlying phase competition mechanisms at boundaries and establish grain boundary commensurability engineering as a critical paradigm for stabilizing the metastable phase and optimizing the macroscopic performance of polycrystalline ferroelectrics.

**Three-Dimensional Boundary Structures in Polycrystalline HZO**

In this work, HZO thin films are grown on $La_{0.67}Sr_{0.33}MnO_3$-buffered $SrTiO_3$ (STO) (001) substrates by PLD and subsequently prepared as freestanding membranes for electron microscopy characterization by selectively wet-etching the $La_{0.67}Sr_{0.33}MnO_3$ layer to release the film from the substrate (**Methods**)[33,34]. X-ray diffraction (XRD) measurements of the as-grown and transferred films (**Extended Data Fig. 1**) identified that the films are predominantly composed of the polar-O phase, exhibiting a preferred <111> crystallographic orientation. Because the O phase nearly adopts a pseudo-cubic structure, in which the lattice parameters along the three orthorhombic axes are similar (**Extended Data Table 1**), its <111>-oriented projection exhibits a slightly distorted hexagonal symmetry (**Extended Data Fig. 2 and Extended Data Table 2**). XRD pole-figure measurements (**Extended Data Fig. 3a**) and four-dimensional scanning transmission electron microscopy (4D-STEM) nanobeam diffraction[31,34,35] reveal two sets of <111> hexagonal diffraction patterns with a 30° rotational offset, confirming the preferred crystallographic orientations of the HZO thin films under epitaxial constraints imposed by the STO substrate (statistics in **Extended Data Fig. 4**).

The preferred orientations give rise to two predominant configurations of *intergranular* grain boundaries with different lattice commensurability (defined in **Extended Data Fig. 2**). The first configuration, with misorientation angles of 0°, 60°, and 120°, exhibits highly coherent lattice structures and favors commensurate grain boundaries. In contrast, the second configuration, with misorientation angles of 30°, 90°, and 150°, inherently introduces geometric frustration, preventing the establishment of short-period atomic registry across the boundary and thereby giving rise to incommensurate grain boundaries. In addition to these *intergranular* grain boundaries described above, *intragranular* ferroelastic domain boundaries can form within a single grain. These domain boundaries retain a coherent lattice (**Extended Data Fig. 5**), in contrast to grain boundaries formed between independently nucleated grains, which typically accommodate lattice mismatch through atomic rearrangement and vacancy incorporation. For simplicity, we hereafter collectively refer to grain boundaries with misorientation angles of 0°, 60°, and 120° as *60° GBs*, and those with misorientation angles of 30°, 90°, and 150° as *30° GBs*. Intragranular ferroelastic domain boundaries are denoted separately as *DBs*.

We employ MEP[32] to investigate boundary structures across multiple grains in these HZO films (**Fig. 2a**), achieving super-resolution and 3D imaging from 4D-STEM convergent beam electron diffraction (CBED) (**Methods**). Ptychographic reconstructions reveal the 3D atomic structure of polycrystalline HZO with various types of boundaries (**Fig. 2b** and **Movie S1**), as well as the coexistence of the dominant polar-O phase and the nonpolar-T phase (**Fig. 2c** and **Supplementary Note S1**). To distinguish these two polymorphs, we measure the local cationic distance (**Extended Data Table 2**). Owing to the distorted pseudo-hexagonal lattice along the <111> orientation, the O phase exhibits alternating lattice spacings along the $[11\bar{2}]$ direction (overlaid in red and blue in **Fig. 2d**, left). The T phase has a nearly ideal hexagonal lattice along <111> orientation, characterized by uniform atomic distances (white in **Fig. 2d**, right). Using cationic spacing alternation as the definitive signature for phase mapping, we examine the bottom interface of the film proximal to the substrate (**Fig. 2e**), and identify continuously curved *30° GBs* with fluctuating inclinations (white curve), straight *60° GBs* without inclinations (green line), and ferroelastic *DBs* (orange line). The polar-O phase dominates the bottom film-substrate interface, where epitaxial strain stabilizes the polar phase[36,37]. In contrast, the top surface of the film (**Fig. 2f**) exhibits the emergence of the nonpolar-T phase proximal to both grain and domain boundaries (white boxes), reflecting partial relaxation of epitaxial strain and surface energy minimization[38,39], which compromises macroscopic ferroelectricity. Notably, the polar-O phase remains entirely preserved even at the top surface exclusively along the commensurate *60° GBs* (green box). This observation reveals an unexpected phenomenon: certain grain boundaries can actively stabilize the polar phase while suppressing the nonpolar phase via localized boundary environment. We postulate that lattice mismatch at the boundaries plays a dominant role in this competition, establishing structural commensurability as a pivotal parameter that can be engineered to control phase stability.

**Polar-Nonpolar Phase Competition at Boundaries**

To validate the lattice commensurability framework, we systematically investigate the 3D atomic structure across three boundary types. At *intragranular DBs*, quantitative depth-resolved analysis of <111>-oriented grains reveals the formation of a "V"-shaped nonpolar T-phase region (**Fig. 3a**). This characteristic V-shaped morphology arises from the interplay between epitaxial strain and its relaxation. Near the bottom film-substrate interface, epitaxial strain partially suppresses T-phase formation, confining it to narrow regions (**Fig. 3b**, orange region). In contrast, away from the substrate, epitaxial strain relaxation drives substantial expansion of the T-phase (**Fig. 3c**, orange region), ultimately reaching approximately 89% T-phase fraction at the top surface (**Fig. 3d,e**). The prevalence of T-phase formation at these boundaries implies that O/T phase interfaces have lower interfacial energy than pure O-phase boundaries, as supported by first-principles calculations (**Extended Data Fig. 6**, **Tables 3** and **4**). To elucidate the microscopic mechanisms driving this competition, we further investigate the same type of ferroelastic *DB* along the <100> orientation (**Extended Data Fig. 7**). This specific viewing direction directly resolves local ferroelastic mismatch arising from the lattice anisotropy of orthorhombic symmetry. Quantitative strain measurements (**Supplementary Note S2**) further reveal that this mismatch is effectively accommodated by the formation of the higher-symmetry nonpolar-T phase, providing a microscopic explanation for its preferential formation at ferroelastic *DBs*.

The polar-nonpolar phase competition at grain boundaries exhibits profound variations, owing to the complex landscape of grain boundaries defined by variable misorientation angles ($\theta$) and inclination angles ($\phi$) (**Extended Data Fig. 2**). For *60° GBs*, the pseudo-hexagonal <111> lattice is characterized by misorientation angle $\theta$ = 0°, 60°, or 120°, with an inclination angle $\phi$ consistently approaching zero, indicating a highly commensurate interface structure (**Fig. 3f**). Depth-resolved atomic-distance maps show the exceptional phase stabilization, where the polar-O phase persists continuously from the bottom (**Fig. 3g**) to the top surface (**Fig. 3h**), with minimal T-phase formation, as further confirmed by additional regions and cross-sectional analysis (**Extended Data Fig. 8** and **Supplementary Note S1**). Notably, this specific grain boundary exhibits periodic atomic vacancies indicated by weaker MEP phase contrast (**Fig. 3i** and **Extended Data Fig. 9**). These ordered vacancies efficiently accommodate the local lattice mismatch between adjacent grains, preserving approximately 95% polar-O phase even at the top surface (**Fig. 3j**). This observation demonstrates an alternative mismatch accommodation mechanism, distinct from the interfacial T-phase buffer observed at *DBs*.

In contrast to the highly commensurate interfaces, the *30° GBs* exhibit a profoundly complex geometry, characterized by a broad range of inclination angles $\phi$ spanning from 0° to 15° (**Extended Data Fig. 2b**). We specifically identify two representative configurations defined by inclination angles of approximately 0° and 15°. For $\phi \approx$ 15°, *30° GBs* introduce severe geometric frustration, compelling most of the top surface to transform into the nonpolar-T phase (**Fig. 3k–m**). Comparing the bottom and top surface (**Extended Data Fig. 10**) reveals a significant twist of the grain-boundary plane. This indicates a depth-dependent evolution of the inclination angle from $\phi \approx$ 15° at the bottom to approximately 0° near the top surface. Driven by this intricately complex 3D grain boundary structure, this incommensurate 30° configuration promotes T-phase formation (approximately 92%) at the top surface and V-shaped morphology (**Fig. 3n,o**) which is also observed in cross-sectional MEP (**Supplementary Note S1**) and becomes more extensive after device cycling (**Supplementary Note S3**). For $\phi \approx$ 0° (**Fig. 3p**), even though localized regions along these optimal zero inclination boundary planes exhibit near coincidence site lattices (NCSLs)[40,41] by approximating a high-index geometric state, the boundary remains poorly commensurate because this excessively large periodicity renders the coincidence matching sites spatially sparse. This geometric constraint dictates excessively large periodic structural units (dashed white boxes in **Fig. 3q,r**). Such weak local mismatch accommodation proves insufficient to fully suppress the phase transition. As a result, the remaining uncompensated strain forces nonpolar-T phase formation asymmetrically along one side of the boundary, leaving approximately 52% polar-O phase at the top surface (**Fig. 3s,t**).

Collectively, these findings indicate that phase competition at boundaries cannot be determined by a single criterion, but instead reflects a delicate balance of competing factors that govern phase stability (**Fig. 3u**). When the boundary is highly commensurate, the intrinsic pseudo-cubic nature of the O-phase introduces local ferroelastic mismatch at *DBs*, which drives degradation into the nonpolar-T phase. In contrast, when the boundary is incommensurate or poorly commensurate, the large geometric frustration disrupts local registry, favoring the formation of a higher-symmetry interfacial T-phase buffer layer to accommodate the severe geometric frustration. Only when an interfacial configuration simultaneously accommodates the pseudo-cubic ferroelastic mismatch while minimizing geometric frustration can the O-phase remain stable across the boundary. To satisfy these conditions, we find that highly commensurate grain boundaries uniquely provide the necessary balance, enabling lattice mismatch accommodation while stabilizing the metastable polar-O phase throughout the boundary plane.

**Thermodynamic Origin of Interfacial Phase Competition**

To uncover the fundamental thermodynamic driving forces behind this commensurability-governed phase competition, we employ Grand Canonical Monte Carlo (GCMC) simulations. Rather than imposing predefined structural constraints, this advanced theoretical framework objectively explores the complex energetic landscapes of these boundaries, revealing the most thermodynamically favored atomic compositions and configurations of grain boundaries (**Methods** and **Supplementary Note S4**). We simulate three distinct grain boundary misorientations ($\theta$ = 60°, 30°, and 38°) for O and T phases. These specifically represent the highly commensurate boundary, the poorly commensurate boundary, and an intermediate level of commensurability, respectively (details in **Methods**). For each specific boundary, the simulated optimal configurations yield atomic-distance maps that are in good agreement with experimental observations. (**Fig. 4a–c** for O phase, **Extended Data Fig. 11** for T phase).

Evaluating the grain-boundary energies ($\gamma$) reveals how critically boundary commensurability dictates phase competition (Details in the **Supplementary Note S4**). At the highly commensurate *60° GBs* (**Fig. 4d**), the simulations identify a single stable configuration characterized by periodic, straight arrays of vacancies (red dashed circles in **Fig. 4a**), consistent with the experimental observations (**Fig. 3g,h**). In contrast, the T phase does not favor this straight array of vacancies. Attempting to enforce this highly ordered vacancy array onto the competing T phase (**Extended Data Fig. 11** and **12**) incurs a substantial energy penalty. Furthermore, the most favorable T-phase configuration remains higher in energy than the most stable O-phase configuration, as shown in **Fig. 4d**. This confirms that the highly commensurate boundary inherently fosters the formation of highly ordered vacancies to stabilize the polar-O phase relative to nonpolar-T phase, rendering O phase thermodynamically robust even near the strain-relaxed free surface.

The pronounced thermodynamic preference provided by *60° GBs* is absent at 38° or 30° *GBs*. Across the chemical potential landscape, both the O- and T-phase *GBs* exhibit a multiplicity of locally stable compositions and structures (**Extended Data Fig. 12**). This structural variety directly reflects the severe geometric frustration and pronounced phase competition inherent to lower-commensurability boundaries. At the 38° *GBs* (**Fig. 4b**), the energy difference between the polar and nonpolar phases ($\Delta\gamma = \gamma_T - \gamma_O$) is notably reduced (**Fig. 4e**). Specifically, under reduced oxygen chemical potentials that closely emulate the practical conditions of the film (orange region in **Fig. 4e**), $\Delta\gamma$ becomes quite small. At the 30° *GBs* (**Fig. 4c**), $\Delta\gamma$ becomes virtually negligible (**Fig. 4f**), indicating the absence of a thermodynamic preference.

To quantify the phase competition across different grain boundaries, we calculate the average grain-boundary energy difference between T and O phases under experimental conditions, where the film typically involves oxygen deficiency. The thermodynamic preference toward polar-O phase conferred by a *GB* decreases from 16.1 to 7.4 and down to 0.8 meV/Å$^2$ as the geometry transitions from the highly commensurate 60° boundary to the poorly commensurate 38° and 30° boundaries, respectively (**Fig. 4g**). This correlation between grain boundary commensurability and thermodynamic phase stability directly rationalizes our atomic-scale observations. At low-commensurability boundaries, the nearly degenerate T- and O-phase grain boundary energies, coupled with the intrinsically higher surface energy of the O phase (**Extended Data Table 4**), inevitably drives the formation of extensive nonpolar T-phase regions observed at the top surface. Ultimately, these GCMC simulations bridge our microscopic structural observations with thermodynamic phase competition. This fundamentally confirms that grain boundary commensurability inherently dictates polar-phase stabilization through highly ordered boundary structure, whereas the geometric frustration dramatically diminishes this thermodynamic preference.

**Enhancing Ferroelectricity Through Grain Boundary Engineering**

Guided by the phase competition insights gained from 3D phase mapping and GCMC simulations, we designed an epitaxial growth strategy to maximize the fraction of highly commensurate *60° GBs*, thereby suppressing the formation of the nonpolar-T phase. The HZO films grown on isotropic $SrTiO_3$ (001) substrate (**Fig. 5a**) mainly form two grain variants with 30° rotational offset and comparable populations along the <111> axis, as verified by nanobeam diffractions (red and orange hexagons in **Fig. 5a, Extended Data Fig. 13a,b**), resulting in a high density of *30° GBs*. To mitigate the formation of *30° GBs* and favor *60° GBs*, we aim to promote the growth of a single grain orientation with near-perfect crystallographic alignment (*i.e.*, 0°, 60°, and 120° misorientation). To achieve this, we employ anisotropic $SrTiO_3$ (110) substrates (**Fig. 5b**), which breaks the degeneracy between grain variants and selects one dominant orientation (orange hexagon in the diffraction pattern), thereby favoring the formation of *60° GBs*. As a result, the polar-O phase dominates in the thin film, as measured by XRD and MEP (**Extended Data Fig. 14, 15**).

To quantify the engineered grain-boundary distribution, we performed large-area 4D-STEM crystallographic mapping[35], which resolves the spatial distribution of grain orientations (**Fig. 5c,d**). In the films grown on the isotropic $SrTiO_3$ (001) substrates, the two dominant grain variants (denoted as <111> #1 and #2) appear in nearly equal fractions (45.1% and 43.9% in **Fig. 5e**), inevitably introducing a high fraction (37.3% in **Fig. 5f**) of the incommensurate *30° GBs*, which could contribute to polar-domain pinning (**Extended Data Fig. 16**). In contrast, for the films grown on the anisotropic $SrTiO_3$ (110) substrates, a single <111> grain variant predominates. This strict orientation control effectively eliminates the *30° GBs* and simultaneously increases the fraction of commensurate *60° GBs* to 66.4% (**Fig. 5f** and **Extended Data Fig. 17**). Collectively, this statistical validation demonstrates that anisotropic epitaxial engineering effectively dictates grain boundary commensurability.

Translating these tailored microstructures into macroscopic performance, the resulting polarization–electric field hysteresis loops (**Fig. 5g**) reveal a substantial enhancement of approximately 60% in remanent polarization for the films grown on $SrTiO_3$ (110), consistent with previous device measurements[42,43]. To evaluate the long-term macroscopic stability, retention measurements were conducted on both types of films (**Fig. 5h**). The films grown on anisotropic $SrTiO_3$ (110) substrates exhibit superior retention properties as compared to those grown on the $SrTiO_3$ (001) substrates, with their remanent polarization projected to remain robust for several months under intrinsic depolarization fields[44]. This study suggests that, even without aggressive film optimization, the films grown on $SrTiO_3$ (110) substrates retain significant polarization, whereas those grown on $SrTiO_3$ (001) substrates show a decay to nearly zero polarization. This highlights grain-boundary engineering as a viable and more accessible route for enhancing device performance.

## Conclusions

This work provides 3D atomic-scale characterization of boundary structures and phase distributions, revealing grain-boundary commensurability as a decisive factor governing polar-phase stability. Our combined investigation using 3D ptychographic imaging and thermodynamic Grand Canonical Monte Carlo simulations uncovers a mechanism in which commensurate grain boundaries thermodynamically stabilize the polar phase. Guided by this mechanism, we successfully engineer grain boundaries to achieve enhanced device performance. These findings shift the paradigm from treating grain boundaries as unavoidable structural defects to leveraging them as tunable functional motifs. This boundary-commensurability engineering strategy offers a general pathway for optimizing phase stability in ferroelectric devices and provides a blueprint for designing interfaces across a broad class of polycrystalline functional materials.

## Methods

### Thin-film Growth

All films were grown using pulsed-laser deposition using a KrF excimer laser (Coherent LPX 305, 248 nm, 25 ns pulse) which was focused on a dense, stoichiometric, ceramic target of the various films noted in the manuscript to deposit material to a prepared substrate. For all growths, the target to substrate distance was 55 mm in an on-axis geometry. Prior to growth, the $SrTiO_3$ substrate was prepared by ultrasonicating for 5 minutes in acetone and isopropyl alcohol. The $SrTiO_3$ substrate was then attached to an Inconel resistive heater using silver paint. $Hf_{0.5}Zr_{0.5}O_2$ and $La_{0.67}Sr_{0.33}MnO_3$ films were deposited at 750 °C (all temperatures measured by a thermocouple within the Inconel heater) in a dynamic oxygen partial pressure of 100 mTorr at a laser fluence of ~1 J/cm$^2$ and a laser repetition rate of 2 Hz. Following deposition of the films, the samples were cooled in a static ~700 Torr oxygen environment at a rate of 10 °C/min.

### X-ray Diffraction

X-ray diffraction $\theta$-$2\theta$ linescans were used to characterize the crystal structure of the films by a Malvern Panalytical X'Pert 3 diffractometer with copper $K_\alpha$ radiation (1.54 Å wavelength). A hybrid, 2-bounce Ge (220) monochromator with 1/2° divergence slit was used in the incidence beam path and the diffracted X-rays were collected by a PIXcel3D detector in single point mode. Linescans were collected at 0.4 seconds per point with a 0.01° step size. XRD pole figures were collected using the HZO {111} reflections to characterize the crystallographic orientation distributions of the as-grown films (**Extended Data Fig. 3**). The as-grown crystallographic orientation analysis of the HZO films is shown by the X-ray diffraction pole figures in **Extended Data Fig. 3**.

### Electrical Measurements

All device-based electrical measurements were taken from parallel-plate, out-of-plane oriented capacitor structures, using a blanket film as the bottom electrode and 12.5-μm-diameter circular contacts as the top electrodes. Following growth, the films were then coated in photoresist (OCG 825 35CS) and cured for 5 minutes on a hot plate at 100 °C. Photolithography was then performed using a shadow mask under ultra-violet (395 nm) light for 8 seconds. Exposed patterns were then developed in a 1:2 solution of Microposit developer concentrate (tetramethylammonium hydroxide) solution and deionized water for 20-30 seconds. Platinum was then deposited at room temperature using DC sputtering in a dynamic 7 mTorr partial pressure of argon. The photoresist was subsequently removed by sonicating in acetone, leaving behind the platinum top contacts.

All ferroelectric measurements were taken using a Radiant Precision Multiferroic Tester. Standard polarization versus electric field hysteresis loops were taken using a double bipolar triangular waveform with a set amplitude (1-10 V) and period (0.1 ms per bipolar loop, or 10 kHz). For retention measurements, the samples are driven to the highest field to saturate the polarization, and the polarization value is then tracked over time. These measurements are performed using a modified positive up and negative down (PUND) sequence, and the details of these measurement protocols are provided in our previous study[45]. Fatigue measurements were performed at room temperature using repeated bipolar pulse bursts at a frequency of 100 kHz and a pulse height of 5 V[45]. After selected cycling intervals, PUND measurements were carried out to extract the remnant switched polarization and monitor its evolution with accumulated switching cycles. No appreciable increase in polarization is observed between the initial measurement and $10^4$ switching cycles, indicating no appreciable wake-up effect under the tested conditions (**Supplementary Note S3**).

**Thin-film Transfer**

The freestanding film was transferred using a procedure consistent with our previous work[27,33,34]. Following growth, the $Hf_{0.5}Zr_{0.5}O_2$ thin film was transferred onto a TEM copper grid for plan-view STEM imaging. To selectively remove the underlying $La_{0.67}Sr_{0.33}MnO_3$ layer without damaging the $Hf_{0.5}Zr_{0.5}O_2$ film, the sample was immersed in an etching solution composed of 8 mg KI, 10 mL HCl, and 100 mL $H_2O$ for approximately 12 hours. This process enabled the release of the thin film from the $SrTiO_3$ substrate. The freestanding $Hf_{0.5}Zr_{0.5}O_2$ film was then collected using a TEM copper grid and dried on a hot plate at 95 °C prior to STEM characterization. Low-magnification transmission electron microscopy (TEM) imaging shows no obvious large-scale tearing or folding (**Extended Data Fig. 18**). The polar-O phase remained stable after transfer release from substrate clamping, suggesting that its stability is maintained through surface and grain boundary energy contributions, in agreement with earlier reports[33].

**4D-STEM Nanobeam Diffraction**

The 4D-STEM datasets were collected using an aberration-corrected ThermoFisher Titan Themis equipped with an electron microscope pixel array detector (EMPAD)[46] operated at 300 kV. The EMPAD features a 128 × 128 pixel array with 1 ms acquisition time, 0.86 ms per frame readout speed, and 1,000,000:1 dynamic range. 4D-STEM nanobeam diffraction employed an electron beam with 1 mrad convergence angle under 285 mm camera length. The experimental condition aligns with methodologies reported in our previous work[47,48]. 4D-STEM nanobeam diffraction enabled the grain orientation mapping across a large field of view in **Fig. 5c–d** and **Extended Data Fig. 4**.

**Multislice Electron Ptychography**

Multislice electron ptychography (MEP) employed a defocused electron probe (approximately 20 nm) with a 25 mrad convergence angle. The electron probe was raster-scanned across the sample, while the convergent beam electron diffraction (CBED) patterns were acquired at each scan position (**Fig. 2a**) using a camera length of 285 mm. The scan step size was set to 0.349 Å, and the electron dose was approximately $10^6$ $e^-$ $Å^{-2}$. The mixed-state[49] MEP method from a customized PtychoShelves[50] package was used to reconstruct the phase information and the corresponding lattice structure from the 4D-STEM dataset. Bayesian optimization method[51] was first applied to determine the accurate defocus value for initial probe generation. Final reconstructions were produced by an agentic framework named "Ptychographic Experiment and Analysis Robot" (PEAR)[52], which leverages large language models and vision language models to automatically explore best reconstruction strategies tailored to each experimental dataset. Ptychographic reconstruction utilized 10 slices (1 nm/slice) and 12 probe modes. The orthogonal probe relaxation (OPR) technique[53] is often used to reduce artifacts caused by probe variation in a single scan, the number of orthogonal modes kept in truncated singular value decomposition controls the number of structural changes allowed at each scan position. Moreover, position correction can refine inaccurate scan positions, and intensity correction accounts for changes in probe intensity. Compared to conventional high-angle annular dark-field (HAADF) imaging, MEP achieves superior lateral resolution (<30 pm) and enables direct visualization of light oxygen atoms (**Extended Data Fig. 19**). Crucially, MEP reconstructs the depth-resolved lattice structure along the electron beam direction, allowing for atomically precise 3D mapping of lattice structure and phase distribution at boundaries.

### Atomic-Distance Analysis from MEP Phase Images

The MEP-reconstructed phase images were analyzed using CalAtom software[54] to determine the precise atomic positions of Hf/Zr atoms via multiple-ellipse fitting, while atomic-distance maps were generated by projecting atomic distances along the specific directions. Owing to the domain-matching epitaxy[20] between the HZO film and the substrate, HZO grains predominantly exhibit a preferred <111> orientation[33]. The <111>-oriented O-phase exhibits alternating atomic distances along $[11\bar{2}]$ (red/blue) (**Fig. 2c**, left), whereas the T-phase exhibits uniform distances (white) (**Fig. 2c**, right). Nearest-neighbor Hf-Hf atomic distances were measured and then projected onto the $[11\bar{2}]$ direction, which represent the interplanar spacings. The detailed phase assignments between O and T-phase from in-plane and cross-sectional direction are discussed in detail in **Supplementary Note S1.** Furthermore, comparisons with projected structural models and simulated ptychographic images of the rhombohedral (R) phase exclude an R-phase assignment for the regions examined in our films, which is also discussed in **Supplementary Note S1.**

### First-principles Theoretical Calculations

Interface, surface, and bulk energies were calculated by first-principles density functional theory (DFT)[55,56] as implemented in the Vienna Ab initio Simulation Package (VASP)[57]. The Perdew–Burke–Ernzerhof (PBE) form[58] of exchange correlation functional within the generalized gradient approximation (GGA)[59] was employed along with the projector augmented wave method for treating core electrons[60] and a plane wave basis set with an energy cutoff of 520 eV. Python Materials Genomics (Pymatgen)[61] and Atomic Simulation Environment (ASE)[62] were used to construct the surface and interface models. Structural optimization and electronic relaxation were carried out using a Γ-centered Monkhorst–Pack[63] k-point sampling grid of 1×4×3 for <111> grain interface structures and 2×1×5 for <100> grain interface structures. Both atomic coordinates and lattice vectors were relaxed until the maximal residual atomic force was less than 0.02 eV/Å$^{-1}$, and the electronic convergence criteria were set to $10^{-5}$ eV. It is worth noting that a fully relaxed interface model tends to transform into a single domain/phase due to the lower bulk free energy and limited supercell size in DFT calculations. To preserve the interface structure, two atomic layers in the middle of each phase were constrained to their bulk configurations during relaxation. Vacancies were introduced by removing Hf atoms at specific sites along the interface. Interface energies were calculated based on the total energy difference between the interface structures and their individual constituent phases as the following: $E_{interface} = (E_{total} - E_a - E_b)/2A_i$, where $E_{total}$ is the energy of the interface structure, $E_a$ and $E_b$ are energies of the two constituent bulk phases, and $A_i$ is the cross-sectional area of the interface. The factor of 2 accounts for the two interfaces present in each periodic interface model. Surface energies were computed by first constructing slabs with specific crystallographic surfaces separated by a vacuum region of 15 Å which mimics the surface condition and minimizes the image interaction due to the periodic boundary condition in the plane-wave DFT calculations. Surface energies were then calculated by the energy difference between the structure with surfaces and the pristine bulk structure: $E_{surface} = (E_s - E_b)/2A_s$, where $E_s$ is the energy of the slab with surfaces, $E_b$ is the energy of the corresponding pristine bulk structure, and $A_s$ is the surface area. The factor of 2 accounts for the two surfaces in each slab model due to the periodic boundary condition in the calculation.

**Grand Canonical Monte Carlo Simulation**

Traditionally, interfacial commensurability in high-symmetry crystals, such as face-centered-cubic metals, is strictly dictated by the crystallographic misorientation angle[64,65]. Within these ideal frameworks, the coincidence site lattice (CSL) theory utilizes the Sigma index (Σ) to rigorously define commensurability based on the spatial density of perfectly matching atomic sites. When extending this classical model to lower-symmetry systems, such as fluorite oxide ferroelectrics, we adopt the near coincidence site lattice (NCSL) to define interfacial commensurability within such pseudo-cubic structures[40,66]. Therefore, retaining Σ index notation from the pseudo-cubic structure, we simulate three representative boundary configurations with progressively decreasing commensurability: 60°, 38°, and 30° misorientations. These specific geometries are analogous to the highly commensurate Σ3, the intermediate Σ7, and the poorly commensurate Σ13 boundaries, respectively. It is crucial to note that our theoretical calculations isolate the thermodynamic effect of misorientation by fixing the inclination angle at 0°. In actual polycrystalline networks, however, for geometries like the 30° boundary, any variations from this optimal 0° inclination introduce structural disorder, eventually degrading the fragile NCSL into an incommensurate state.

In the GCMC framework, we fix the chemical potential $\mu$, volume $V$, and temperature $T$, and let the number of atoms $N$ and energy $U$ evolve[67–69]. We employ a pretrained universal machine-learning interatomic potential (uMLIP), Equiformer V2[70], to evaluate the total energy of the system $U$, which provides the accuracy close to DFT and improves the efficiency of structure evolution significantly. The free energy of the system under various chemical potentials is defined as follows:

$$F = U - \left(\Delta\mu_{\mathrm{Hf}} + \mu_{\mathrm{Hf}}^{0}\right)N_{\mathrm{Hf}} - (\Delta\mu_{\mathrm{O}} + \mu_{\mathrm{O}}^{0})N_{\mathrm{O}} \quad (1)$$

where $U$ is the total energy of the grain boundary structures from the uMLIP calculation, and $\Delta\mu = \mu - \mu^0$ is the relative chemical potential, and $N$ is the number of atoms of each element. Since the grain boundary is in equilibrium with the bulk $HfO_2$, the chemical potential satisfies the relationship:

$$\Delta\mu_{\mathrm{Hf}} + 2 \cdot \Delta\mu_{\mathrm{O}} = \Delta H_{\mathrm{f}}(\mathrm{HfO_2}) \quad (2)$$

where $\Delta H_f(HfO_2)$ is the formation energy of bulk $HfO_2$ for either the orthorhombic or tetragonal phase. The formation energy is calculated by:

$$\Delta H_{\mathrm{f}}(\mathrm{HfO_2}) \;=\; E_{\mathrm{bulk}}(\mathrm{HfO_2})(\mathrm{f.u.}) - E(\mathrm{Hf}) - E(\mathrm{O_2}) \quad (3)$$

where $E_{bulk}(HfO_2)(f.\,u.)$ is the energy of the orthorhombic or tetragonal $HfO_2$ per formula unit, and we estimate the chemical potential $\mu_{Hf}{}^0 = E(\mathrm{Hf})$ and $\mu_O{}^0 = \mathrm{E}(O_2)/2$, as the energies of Hf and half of $O_2$ bulk single crystal and molecule, respectively. By plugging Equations (2) and (3) into (1), we can reduce it to:

$$F = U - N_{\mathrm{O}}/2 \cdot E(\mathrm{HfO_2}) - (N_{\mathrm{Hf}} - N_{\mathrm{O}}/2) \cdot \left(\Delta\mu_{\mathrm{Hf}} + \mu_{\mathrm{Hf}}^{0}\right) \quad (4)$$

or

$$F = U - N_{\mathrm{Hf}} \cdot E(\mathrm{HfO_2}) - (N_{\mathrm{O}} - 2 \cdot N_{\mathrm{Hf}}) \cdot (\Delta\mu_{\mathrm{O}} + \mu_{\mathrm{O}}^{0}) \quad (5)$$

Each grain boundary structure corresponds to a line in the free energy phase space, either ($F$, $\Delta\mu_{Hf}$) or ($F$, $\Delta\mu_O$), as shown in **Extended Data Fig. 12**. As seen, the free energy of each grain boundary varies continuously with the chemical potential. At any given value of $\Delta\mu_O$, the grain boundary structure with the lowest free energy is thermodynamically favored. The intersections of the lines form the free energy convex hull: only the segments forming the lower envelope of the hull correspond to thermodynamically stable grain boundary structures at a specific chemical potential range. Grain boundary structures that lie above the hull are metastable. For the 60° and 30° boundaries exhibiting multiple stable structures in the phase diagram, we selected the structures that best agree with the experimental conditions and observations to construct periodic boundary condition supercells for grain boundary energy calculation, as described in the **Supplementary**

**Note S4**. For the 38° boundary, we adopted the thermodynamically most stable O- and T-phase configurations at lower oxygen chemical potential due to the oxygen-deficient HZO films. To obtain the grain boundary energy, $\gamma$, we constructed supercells with periodic boundary conditions containing two grain boundary interfaces. The value of $\gamma$ was calculated by subtracting the bulk energy and dividing by two to account for the two interfaces, as detailed in **Supplementary Note S4**.

The averaged grain boundary energy difference between T- and O-phase in **Fig. 4g** is calculated as

$$\overline{\Delta\gamma} = \frac{1}{b-a}\int_a^b (\gamma_{\mathrm{T}} - \gamma_{\mathrm{O}})\mathrm{d}\Delta\mu_{\mathrm{O}} \tag{6}$$

where we select an oxygen-deficient regime, ranging from $\Delta H_f(HfO_2)/2$ = -6 eV (a) to -2 eV (b), to mimic the typical oxygen-deficient thin films. The lower bound is determined by the chemical potential stability window of $HfO_2$, while the upper bound corresponds to the threshold below which the oxygen-deficient $HfO_2$ can be energetically favored (details in **Supplementary Note S4**).

**Acknowledgements**

C.S., X.L., R.R., L.W.M., and Y.H. acknowledge the support from the National Science Foundation under grants DMR-2329111 and CMMI-2239545 and the Welch Foundation (C-2065). X.L. acknowledges support from the Rice Advanced Materials Institute (RAMI) at Rice University as a RAMI Postdoctoral Fellow. J.S. acknowledges the support of the U.S. Department of Energy, Office of Science, Office of Basic Energy Sciences, under Award Number DE-SC-0012375 for the development of novel ferroelectric materials. L.W.M. acknowledges additional support from the Intel COFEEE program. C.S., X.L. and Y.H. acknowledge the Electron Microscopy Center, Rice University. K.Z. and X.Q. acknowledge the support from the National Science Foundation under award CMMI-2226908 and DMR-2103842. Portions of this research were conducted with the advanced computing resources provided by Texas A&M High Performance Research Computing. Y. J. acknowledges support from the U.S. DOE Office of Science-Basic Energy Sciences, under Contract No. DEAC02-06CH11357. X.H. and A.M.R. acknowledge support by the Office of Naval Research, under grant number N00014-24-1-2500, for the study of the composition, structures, and energetics of $HfO_2$ grain boundaries. Computational support was provided by the National Energy Research Scientific Computing Center (NERSC), a U.S. Department of Energy, Office of Science User Facility located at Lawrence Berkeley National Laboratory, operated under Contract No. DE-AC02-05CH11231.

**Author contributions**

Conceptualization: C.S., X.L. and Y.H. Methodology: C.S., X.H., K.Z., J.S., Y.J., R.R., X.Q., L.W.M., A.M.R., X.L. and Y.H. Investigation: C.S., X.H., X.L. and Y.H. Visualization: C.S., X.H., Y.J., A.M.R., X.L. and Y.H. Supervision: R.R., X.Q., L.W.M., A.M.R. and Y.H. Writing – original draft: C.S., X.H., X.L. and Y.H. Writing – review & editing: C.S., X.H., K.Z., J.S., Y.J., R.R., X.Q., L.W.M., A.M.R., X.L. and Y.H.

**Competing interests:** There are no competing interests to declare.

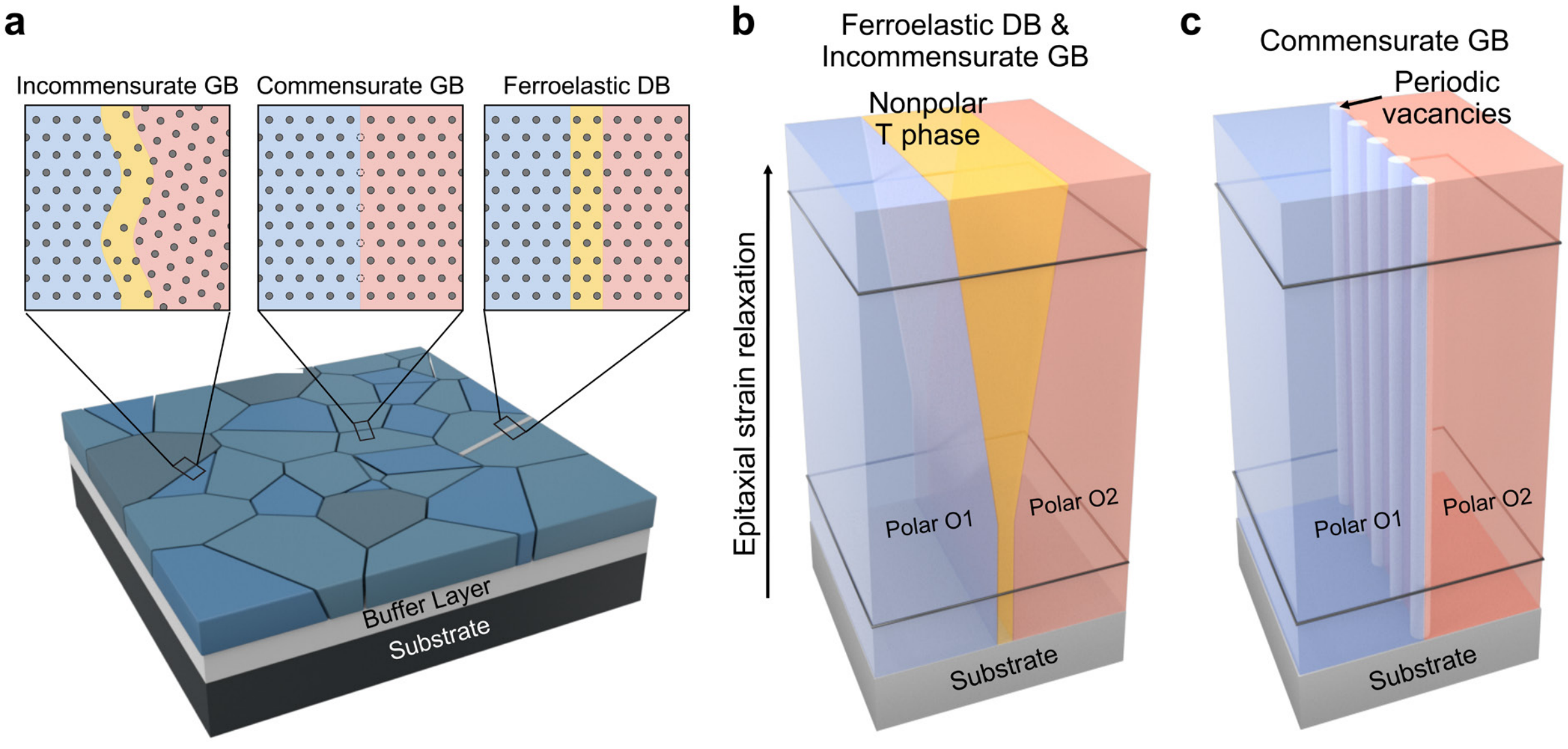

**Fig. 1 | Three-dimensional structural inhomogeneity and phase competition at HZO boundaries**. **a**, Schematic of a polycrystalline HZO thin film. The insets highlight the three predominant interface structures: geometrically frustrated incommensurate grain boundaries (GBs), highly ordered commensurate GBs, and ferroelastic domain boundaries (DBs). Red and blue backgrounds denote two polar-O phase grains or domains. Yellow background represents interfacial nonpolar-T phase region. The dashed circles within the commensurate GBs illustrate periodic vacancies at the boundary plane. **b**, 3D schematic illustrating the three-dimensional phase distribution at typical incommensurate GBs and ferroelastic DBs (collectively denoted as "other boundaries"). As epitaxial strain relaxes along the out-of-plane direction, uncompensated lattice mismatch drives the preferential formation of the higher-symmetry nonpolar T phase at boundaries between polar-O phase (O1 and O2), resulting in a characteristic V-shaped phase distribution. **c**, 3D schematic of a highly ordered commensurate GB, which exhibits ordered periodic vacancies to structurally accommodate lattice mismatch. This alternative relaxation mechanism suppresses T-phase formation, robustly preserving the polar-O phase throughout the entire film.

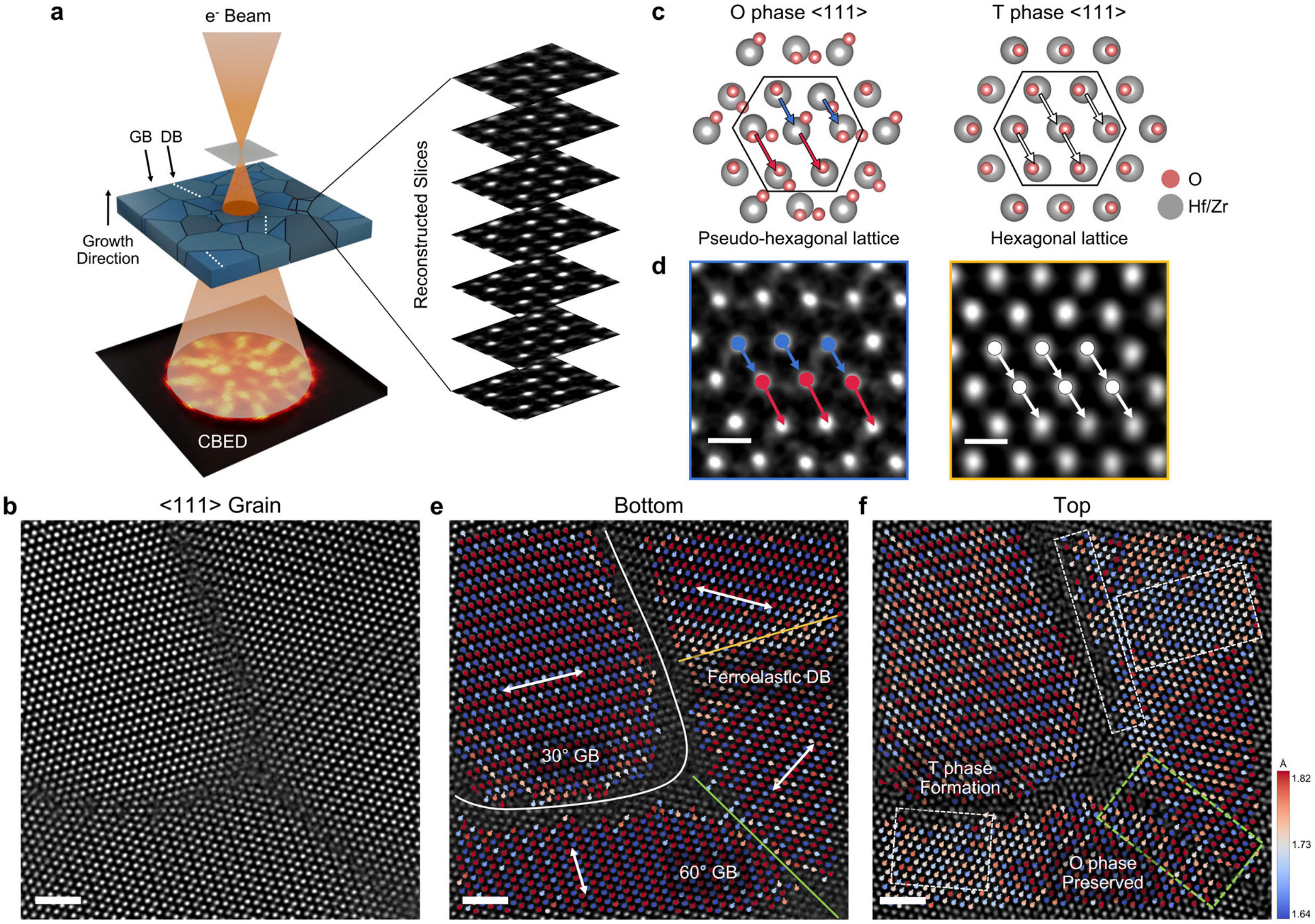

**Fig. 2 | Atomic-scale 3D mapping of interfacial phase distribution. a,** Schematic of the multislice electron ptychography (MEP) experimental setup. A defocused electron beam scans across the polycrystalline HZO thin film, capturing convergent beam electron diffraction (CBED) patterns to extract 3D structural information across both grain boundaries (GBs) and domain boundaries (DBs). **b,** Ptychographic phase image of the polycrystalline HZO film with diverse boundaries. **c,d**, Schematic (**c**) and ptychographic phase images (**d**) of the nonpolar-T and polar-O phases along the <111> zone axis. The phases are distinguished by their different projected atomic distances: the pseudo-hexagonal O phase exhibits alternating spacings (labeled by red and blue arrows), whereas the higher-symmetry T phase displays uniform distances (white arrows). **e,f,** Atomic-distance maps along the $[11\bar{2}]$ direction at the bottom (**e**) and top (**f**) of the film. White arrows denote the $(1\bar{1}0)$ lattice plane normal to the $[11\bar{2}]$ direction. Solid lines represent incommensurate *30° GBs* (white), commensurate *60° GBs* (green), and ferroelastic *DBs* (yellow). At the bottom interface near the substrate (**e**), the polar-O phase predominates. However, at the top surface (**f**), the nonpolar-T phase emerges surrounding both the *30° GBs* and the *DBs* (dashed white boxes). In contrast, the polar-O phase remains preserved exclusively surrounding the *60° GBs* (dashed green box). Colorbar unit: Å. Scale bars: 2 Å in **d**; 1 nm in **b,e,f**.

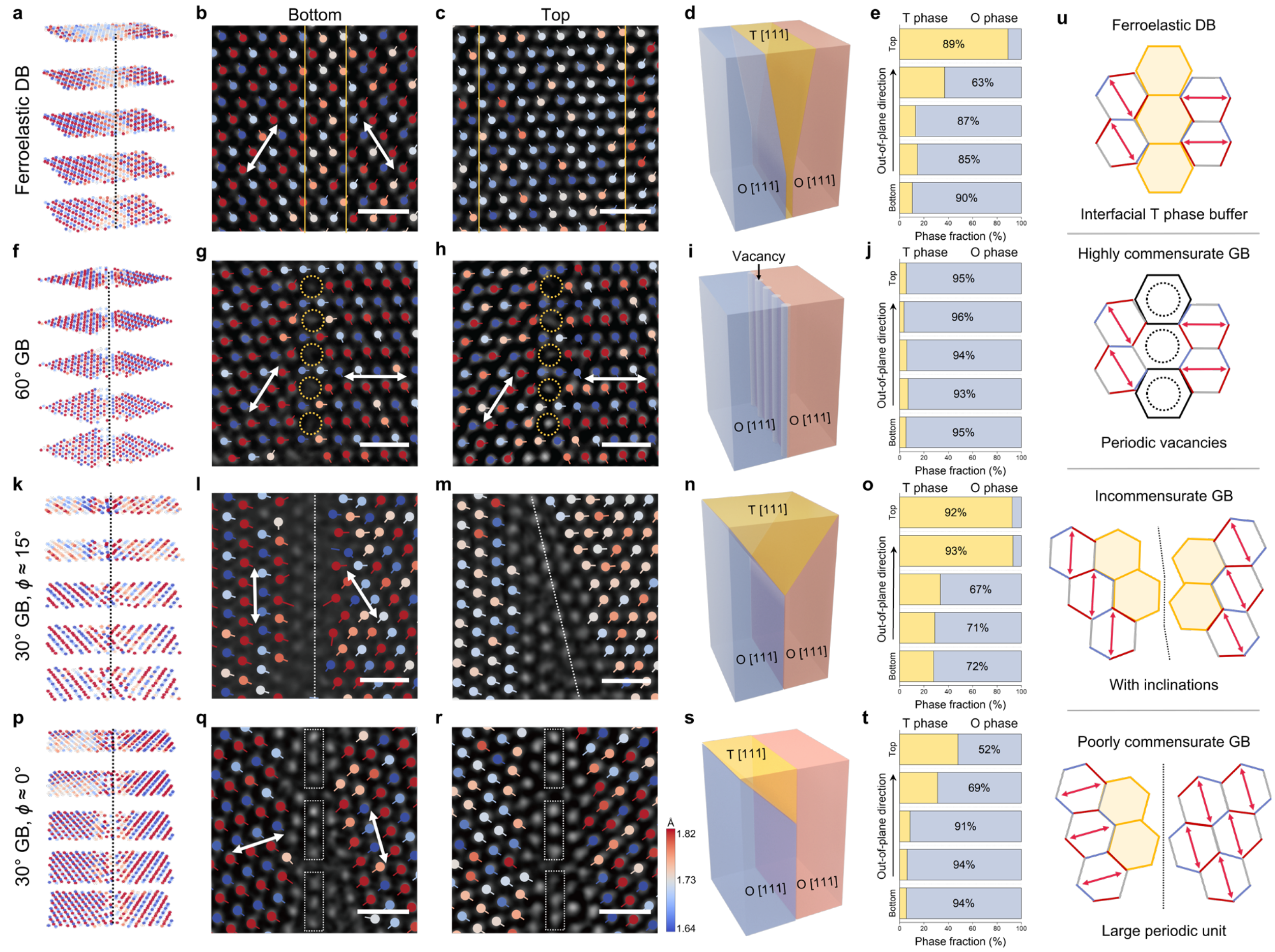

**Fig. 3 | O- and T-phase competition at boundary structures. a,** Depth-resolved 3D atomic-distance maps along $[11\bar{2}]$ direction of ferroelastic domain boundary (DB). **b,c,** Atomic structures overlaid with atomic-distance maps at bottom (**b**) and top (**c**) layers. Yellow lines mark interfacial T-phase regions. White arrows represent polarization direction in each grain. **d,e,** 3D schematic (**d**) and quantitative phase competition analysis (**e**) at ferroelastic DB. **f,** Depth-resolved 3D atomic-distance maps of commensurate 60° grain boundary (GB). **g,h,** Atomic structures overlaid with atomic-distance maps at bottom (**g**) and top (**h**) layers. Yellow dashed circles highlight the periodic atomic vacancies. **i,j,** 3D schematic (**i**) and quantitative phase competition analysis (**j**). **k,** Depth-resolved 3D atomic-distance maps of 30° GB with inclination angle of approximately 15°. **l,m,** Atomic structures overlaid with atomic-distance maps at bottom (**l**) and top (**m**) layers. **n,o,** 3D schematic (**n**) and quantitative phase competition analysis (**o**). **p,** Depth-resolved 3D atomic-distance maps of 30° GB with inclination angle of approximately 0°. **q,r,** Atomic structures overlaid with atomic-distance maps at bottom (**q**) and top (**r**) layers. Dashed white boxes highlight localized near-coincidence site lattices (NCSLs). **s,t,** 3D schematic (**s**) and quantitative phase competition analysis (**t**). **u,** Schematic illustrates the typical interfacial structures and their corresponding lattice mismatch accommodation mechanisms. Hexagons outlined in red/blue denote long/short atomic distance of polar-O

phase. Red arrows indicate the projected polarization direction. Yellow hexagons represent interfacial T-phase region. The distinction between ferroelastic DBs and commensurate GBs is discussed in **Supplementary Note S2**. Scale bars: 5 Å

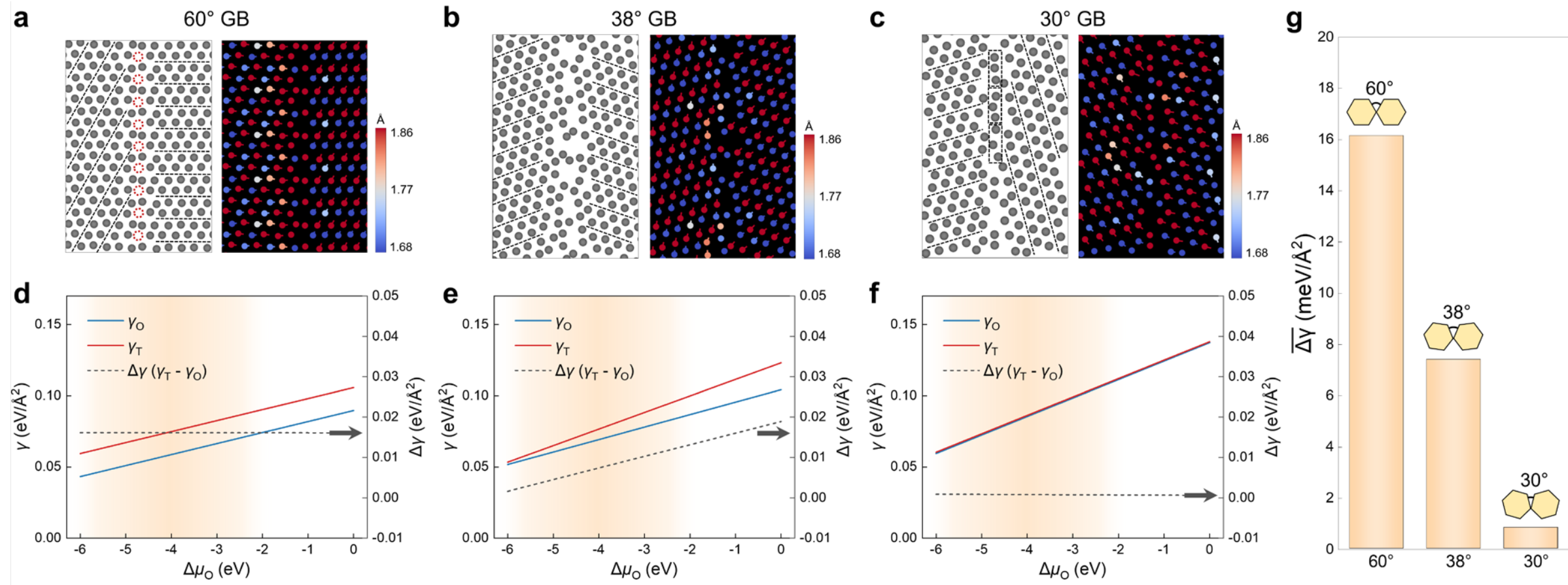

**Fig. 4 | Thermodynamic origins of commensurability-driven phase competition. a–c,** Grand canonical Monte Carlo (GCMC)-simulated atomic structure of 60° (**a**), 38° (**b**), and 30° (**c**) O-phase grain boundaries (GBs), alongside corresponding atomic-distance maps along $[11\bar{2}]$ direction. For clarity, only Hf atoms are displayed. Dashed red circles in **a** highlight the straight columns of periodic Hf vacancies unique to the highly commensurate 60° boundary. The red and blue contrast near the GB planes reflects variations in projected atomic distances associated with local structural disorder. **d–f,** Grain boundary energies ($\gamma$) of 60° (**d**), 38° (**e**) and 30° (**f**) grain boundaries as functions of the oxygen chemical potential ($\Delta\mu_O$) with different competing phases. Dashed black lines indicate the energy difference between two phases ($\Delta\gamma = \gamma_T - \gamma_O$) on the right-side $y$-axis. The orange regions highlight the slightly oxygen-deficient regime characteristic of experimental HZO films. At the commensurate 60° boundary with ordered vacancies (**d**), the polar-O phase is energetically preferred over the T-phase due to its lower GB energy. Conversely, at the geometrically frustrated 38° (**e**) and 30° (**f**) boundaries, this energetic preference collapses, resulting in intense phase competition. **g,** Averaged grain boundary energy difference ($\overline{\Delta\gamma}$) between nonpolar-T and polar-O phase for different grain boundaries across the experimental oxygen-deficient regime ($\Delta\mu_O$ from -6 to -2 eV).

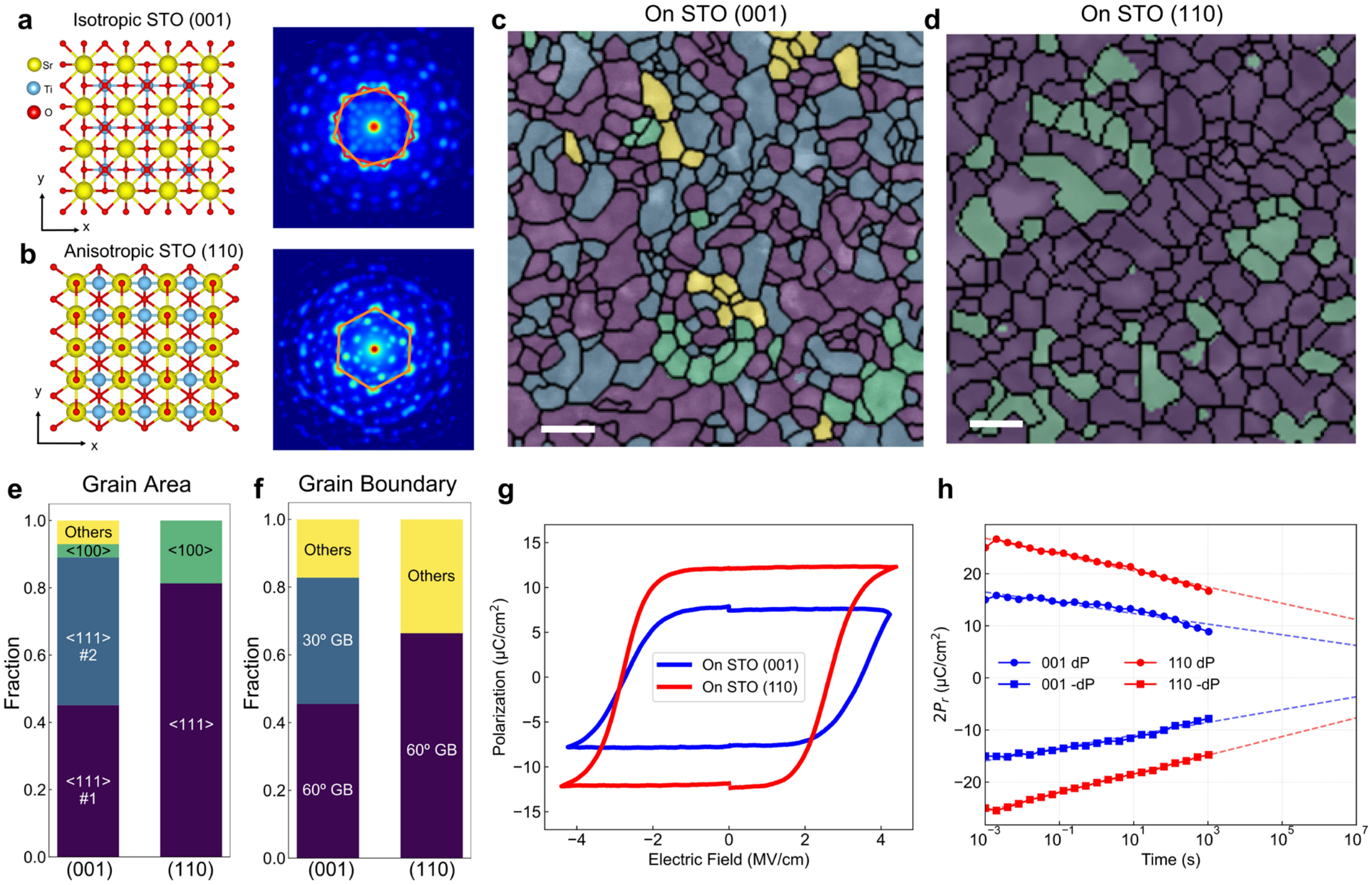

**Fig. 5 | Macroscopic ferroelectric enhancement via grain boundary commensurability engineering. a,b,** Atomic schematics of isotropic STO (001) and anisotropic STO (110) substrates (left) alongside the corresponding 4D-STEM nanobeam electron diffraction patterns of the HZO films (right). The isotropic STO (001) substrate (**a**) exhibits two distinct <111> orientation sets with a 30° rotational offset, evidenced by the two sets of pseudo-hexagonal diffraction spots (red and orange). The anisotropic STO (110) substrate (**b**) breaks this crystallographic degeneracy, promoting the preferential formation of one dominant <111> set (orange). **c,d,** Large-scale mapping of crystallographic orientations and grain boundary networks by unsupervised machine learning for films grown on STO (001) (**c**) and STO (110) (**d**) substrates. Two <111> sets (purple and blue) with a 30° rotational offset coexist with <100> grains (green). Yellow indicates grains with unclassified orientations due to severe mistilt. **e,f,** Statistical quantification of the total grain area (**e**) and grain boundary length (**f**). The two sets on the isotropic substrate induce a high fraction (approximately 37.3%) of geometrically frustrated 30° boundaries. By promoting a single <111> set, the anisotropic substrate eliminates these 30° boundaries and maximizes the fraction of highly commensurate 60° boundaries to approximately 66.4%. **g,h,** *P*–*E* hysteresis loops (**g**) and ferroelectric retention as a function of elapsed time after polarization writing (**h**) of HZO films grown on the two substrates, showing approximately 60% enhancement of remanent polarization with superior retention projected to remain robust for several months. Scale bars: 20 nm.